\documentclass[prb,twocolumn,showpacs,preprintnumbers,amsmath,amssymb]{revtex4}
\usepackage{graphicx}% Include figure files
\usepackage{dcolumn}% Align table columns on decimal point
\usepackage{bm}% bold math

\begin{document}

\def\ket#1{\langle#1\mid}
\def\bra#1{\mid#1\rangle}

\title{Block-block entanglement and quantum phase transitions
in one-dimensional extended Hubbard model}

\author{Shu-Sa Deng} \author{Shi-Jian Gu}  \author{Hai-Qing Lin}
\affiliation{Department of Physics and Institute of Theoretical Physics, The
Chinese University of Hong Kong, Hong Kong, China}

\begin{abstract}
In this paper, we study block-block entanglement in the ground state of
one-dimensional extended Hubbard model. Our results show that the phase
diagram derived from the block-block entanglement manifests richer
structure than that of the local (single site) entanglement because it
comprises nonlocal correlation. Besides phases characterized by the
charge-density-wave, the spin-density-wave, and phase-separation, which can
be sketched out by the local entanglement, singlet superconductivity phase
could be identified on the contour map of the block-block entanglement.
Scaling analysis shows that ${\rm log}_2(l)$ behavior of the block-block
entanglement may exist in both non-critical and the critical regions, while
some local extremum are induced by the finite-size effect. We also study
the block-block entanglement defined in the momentum space and discuss its
relation to the phase transition from singlet superconducting state to the
charge-density-wave state.
\end{abstract}


\date{\today}
\maketitle

\section{introduction}

Entanglement, as one of the most intriguing feature of quantum mechanics
\cite{ABinstein35} and a crucial resource in quantum information theory
\cite{SeeForExample,MANielsenb}, has been a subject of much study in recent
years. For the purpose of practical application, to realize quantum information
processing, such as quantum state transfer \cite{YLi05,JZhang05}, based on the
condensed matter physics is of great importance. Such potential prospect in the
application now has motivated many theoretical investigations on the
entanglement in spin and fermionic systems
\cite{KMOConnor2001,PZandardi2000,LFSantos2003,XWang2001PLA,SJGuPRAJ1J2}. On
the other hand, quantum phase transitions (QPTs), which are characterized by
change in the properties of the ground state of a many-body system, are
generally driven by quantum fluctuation at zero temperature. One therefore
expects that the entanglement, as a term of pure quantum correlation, should
closely relate to QPTs. Indeed, some observed relationships
\cite{AOsterloh2002,SJGuXXZ,JVidal04,LAWu04,MFYang04,YChen04,FVerstraete04,KAudenaert02,GVidal2003,VEKorepin04,LViola04}
between the entanglement and QPTs \cite{Sachdev} now have not only deepened our
understanding on the QPTs, but also strengthened the connection between the
quantum information theory and condensed matter physics. For example, Osterloh
{\it et al.,} \cite{AOsterloh2002} reported that the pairwise entanglement of
two nearest neighbors shows scaling behavior in the vicinity of quantum phase
transition point of the transverse-field Ising model. For other models, such as
the XXZ model \cite{SJGuXXZ}, spin model with mutual exchanges \cite{JVidal04},
etc., the pairwise entanglement also manifests various interesting properties,
such as being maximum at transition point, exhibits singularity, and shows
scaling behavior.

Besides the pairwise entanglement between two sites, for a non-degenerate
ground state, the block-block entanglement \cite{KAudenaert02,GVidal2003} also
provides a good quantity to describe the pure quantum correlation at zero
temperature. The scaling property of the block-block entanglement establishes
an interesting connection between concepts of quantum information and quantum
field theory \cite{VEKorepin04}. Moreover, unlike the concurrence \cite{Hill},
a measurement of the entanglement of formation, which is computable only for
spin-1/2 system, the block-block entanglement can be generalized to high-spin
and fermionic systems. There are some works which studied the entanglement in
fermionic systems
\cite{JSchliemann_PRB_63_085311,PZanardi_JPA_35_7947,PZanardi_PRA_65_042101}.
However, the investigations on the entanglement in relation to QPTs are still
on an early stage \cite{SJGuPRL,JWang03,AAnfossi,SSDeng05,DLarsson05}.

In the extended Hubbard model, it has been shown that the global phase diagram
can be sketched out by the contour map of a quantity called the local
entanglement \cite{PZanardi_PRA_65_042101,SJGuPRL}. However, the local
entanglement fails to identify phases which are related to the
off-diagonal-long-range order, such as the superconducting phase. In this
paper, we extend our previous investigation on the local entanglement in the
extended Hubbard model to the block-block entanglement. We show that the
block-block entanglement with block size larger than one can provide more
useful information than the local entanglement since it comprises the nonlocal
correlation in its expression. The paper is organized as follows. In Sec.
\ref{sec:model}, we first briefly review some basic knowledge of the extended
Hubbard model and then introduce the definition of the block-block
entanglement. In Sec. \ref{sec:level3}, we address the problem of the global
phase diagram on the $U-V$ plane by introducing the contour map of the
block-block entanglement. In Sec. \ref{sec:level4}, we study the scaling
behavior of the block-block entanglement by changing the system size. One of
the interesting results obtained in this work is that the $\log_2(l)$ behavior
($l$ is the block size) of the block-block entanglement also shows in
noncritical region. In Sec. \ref{sec:level5}, we study the finite-size effort
and clarify some unaccustomed behaviors of the block-block entanglement. A
simple expression of the local entanglement is obtained. In Sec.
\ref{sec:further}, we try other measure of the block-block entanglement in
order to explore the phase boundary which may not be identified by the extremum
of the local entanglement. Finally, a brief summary is given in Sec.
\ref{sec:sum}.

\section{The model and formulism}
\label{sec:model}

The one-dimensional extended Hubbard model is defined by the Hamiltonian
\begin{eqnarray}\label{eq:Hamiltonian}
H=-\sum_{\sigma,j,\delta}c^\dagger_{j,\sigma}c_{j+\delta,
\sigma}+U \sum_j n_{j
 \uparrow}n_{j \downarrow}+V\sum_j n_j n_{j+1},
\end{eqnarray}
where $\sigma=\uparrow,\downarrow;\,j=1,\dots, L; \delta=\pm 1$,
$c^\dagger_{j, \sigma}$ and $c_{j, \sigma}$ are creation and annihilation
operators at site $j$, $U$ and $V$ define the on-site and nearest-neighbor
Coulomb interactions, respectively. On each site the local states have four
possible configurations, denoted by
\begin{eqnarray}\label{eq:Localbasis}
|\phi(l=1,2,3,4)\rangle=|0\rangle,\;|\uparrow\rangle,\;
|\downarrow\rangle,\;|\uparrow\downarrow\rangle.
\end{eqnarray}
The Hilbert space associated with $L$-site system, known as the Fock space
${\cal H}_F(L)$, is spanned by $4^L$ basis vectors $|j_1,\dots,
j_L\rangle=\prod_{j=1}^L|\phi_j(l_j)\rangle$. Any state in such a system
can be expressed as a superposition of these basis. If we choose periodic
boundary condition for $L=4n+2$ and anti-periodic boundary condition for
$L=4n$, where $n$ is an arbitrary integer, the ground state is
non-degenerate \cite{GSTian}. Therefore, considering the reduced density
matrix of a block of $l$ successive site of the ground state $\rho_l={\rm
tr}_r |\Psi\rangle\langle\Psi|$, the von Neumann entropy $E_v(l)$, i.e.
\begin{eqnarray}
E_v(l)=-{\rm tr}[\rho_l\log_2(\rho_l)]
\end{eqnarray}
measures the entanglement between the $l$ sites and the rest $L-l$ sites of the
system, as shown in the following figure for 10-site system with $l=4$.

\begin{figure}[h]
\includegraphics[width=7.5cm]{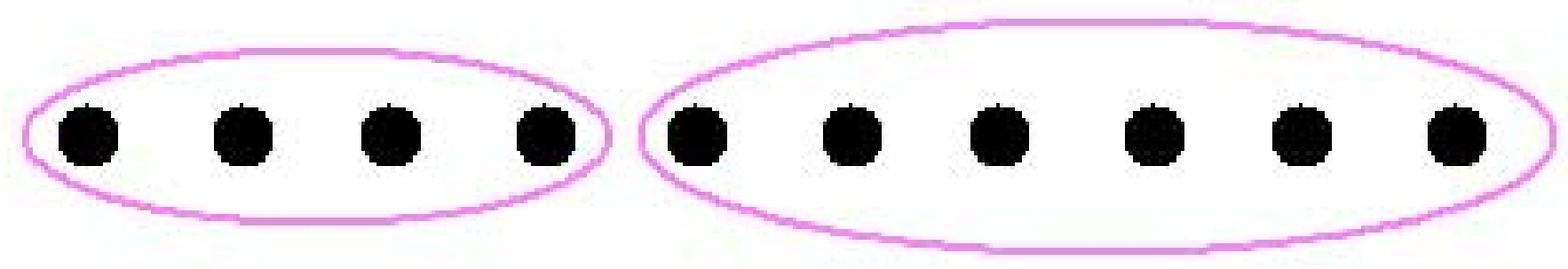}
\end{figure}

A simple case of the block-block entanglement is when only one site is taken
into account. Then the reduced density matrix can be written into a simple form
\cite{PZanardi_PRA_65_042101,SJGuPRL},
\begin{eqnarray}
\rho_1= z\bra{0}\ket{0} + u^+\bra{\uparrow}\ket{\uparrow}
         + u^- \bra{\downarrow}\ket{\downarrow}
          + w \bra{\uparrow\downarrow}\ket{\uparrow\downarrow} ,
\end{eqnarray}
in which
\begin{eqnarray}
w&=&\langle n_{\uparrow}n_{\downarrow}\rangle
 = {\rm Tr}(n_{ \uparrow}n_{ \downarrow}\rho_1), \nonumber\\
  u^+&=&\langle n_\uparrow\rangle - w,  \;\;
   u^-=\langle n_\downarrow\rangle - w, \nonumber\\
    z&=&1 - u^+ - u^- -w
    = 1 - \langle n_\uparrow\rangle - \langle n_\downarrow\rangle + w.
\end{eqnarray}
Here $\langle n_\uparrow\rangle$ and $\langle n_\downarrow\rangle$ are electron
densities with spin up and spin down respectively. The corresponding von
Neumann entropy then takes the form
\begin{eqnarray}
E_v=-z\log_2z-u^+\log_2u^+ -u^-\log_2u^- -w\log_2w.  \nonumber
\end{eqnarray}
If $l>1$, the reduced density matrix can not be written into a simple
expression. However, since the Hamiltonian (\ref{eq:Hamiltonian}) has
U(1)$\times$SU(2) symmetry: $c_{j,\sigma}\rightarrow
e^{i\theta}c_{j,\sigma};\; c_{j,\sigma}\rightarrow
U_{\sigma\delta}c_{j,\delta}$, which manifests the charge conservation and
invariance under spin rotation $U_{\sigma\delta}$. Therefore, any
eigenstate of the Hamiltonian (\ref{eq:Hamiltonian}) can be both the
eigenstate of $z$ component of total spins and of particle number. This
fact leads to that, for arbitrary block size $l$, there is no coherent
superposition of local state with different value of $S^z(l)=\sum_j
(n_{j,\uparrow}-n_{j,\downarrow})$ and
$N(l)=\sum_j(n_{j,\uparrow}+n_{j,\downarrow})$ in the reduced density
matrix. That is, the reduced density matrix must have the diagonal form
classified by both $S^z(l)$ and $N(l)$. In this paper, we apply exact
diagonalization technique to get the ground state and then exactly
diagonalize the reduced density matrix to calculate the block-block
entanglement.

\section{Entanglement and the phase diagram}
\label{sec:level3}

\begin{figure*}
\includegraphics[width=15cm]{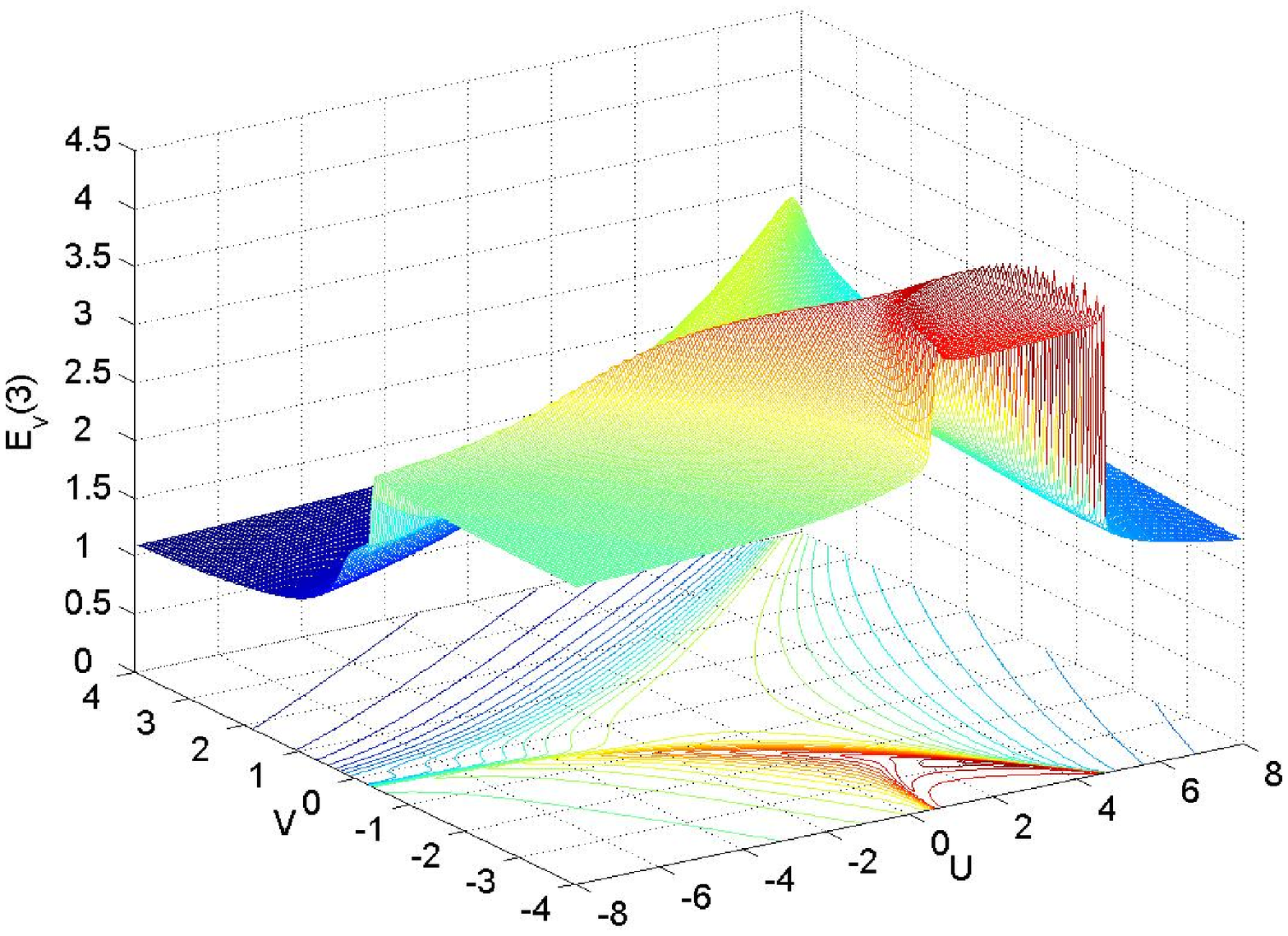}
\caption{The block-block entanglement and its contour map changes with
parameter $U$ and $V$. Here $L=8, l=3$. \label{fig:3d2} }
\end{figure*}

The extended Hubbard model is a prototype model in condensed matter theory for
it exhibits a rich phase diagram \cite{Emery,Solyom,Lin} where various quantum
phase transitions occur among symmetry broken states. The corresponding
symmetry broken states typically include the charge-density-wave (CDW), the
spin-density-wave (SDW), phase separation (PS), singlet (SS) and triplet
superconducting phase(TS), and bond-order wave (BOW)
\cite{MNakamura00,PSengupta02,EJeckelmann02}. Many efforts have been made to
obtain the phase diagram of the extended Hubbard model at different band
fillings on the U-V plane \cite{Lin,MNakamura00,PSengupta02,EJeckelmann02}. In
our previous work on the local entanglement \cite{SJGuPRL}, it is remarkable to
see that the skeleton of the phase diagram of the extended Hubbard model can be
directly obtained from the contour map of the local entanglement. This is by no
means trivial. In the conventional approach to obtain the phase diagram of the
extended Hubbard model, one has to study behaviors of different order
parameters in different regions, either by comparing the ground-state energy or
the critical exponent of correlation functions associated with broken symmetry.
While using a single quantity, $E_v$, the global picture of the system at zero
temperature can be observed. This is not a coincident, for the non-vanishing
order parameter means the existence of non-local corresponding correlation,
whose quantum part is just the entanglement, so the competitions between
different orders may lead to changes in the entanglement. When a QPT occurs,
entanglement will also behave distinctively in different phases. Therefore,
this result reflects the underlying relation between the entanglement and QPTs
beyond the superposition principle of quantum mechanics. However, there are
some limitations. For example, the local entanglement can not be used to
identify superconducting phases, due to the fact that the broken symmetry is
associated with off-diagonal-long-range-order, whereas the local reduced
density matrix is diagonal. To identify such phase is a challenge, and it is
one of our motivations to include the off-diagonal correlation in the study of
the entanglement and QPTs.

In Fig. \ref{fig:3d2}, we show a three-dimensional diagram, as well as its
contour map of the block-block entanglement for 8 sites system with block
size $l=3$. Obviously the structure of the contour map is richer than that
of the local entanglement \cite{SJGuPRL}, especially at the region near
$U<0, -1<V<0$. The contour lines should be similar in the same phase
region, because the entanglement should behave in a similar way in the same
phase. So the distinct change of the contour lines suggests the existence
of a new phase. Compare this contour map with the phase diagram of the
extended Hubbard model at half-filling, this region corresponds to SS
phase. For the transition from CDW to SDW, the block-block entanglement,
including the local entanglement\cite{SJGuPRL} exhibits maximal value. This
fact can be understood by noting that, at the QPT point from CDW to SDW,
the weights for different component in the reduced-density matrix are
closer due to the higher symmetry in the critical region. Take the local
entanglement as an example, we have equally valued $w, u^+, u^-$, and $z$
at the transition point. From this point of view, the extremum behavior of
the entanglement at the transition point manifests that the QPTs are
induced by symmetry broken \cite{SJGuPRL}. For the transition from SS to
CDW, the contour lines have an inflexion, which suggests the existence of
other types of phase transitions.

\section{\label{sec:level4}Entanglement change with block size}

\begin{figure}
\includegraphics[width=7.5cm]{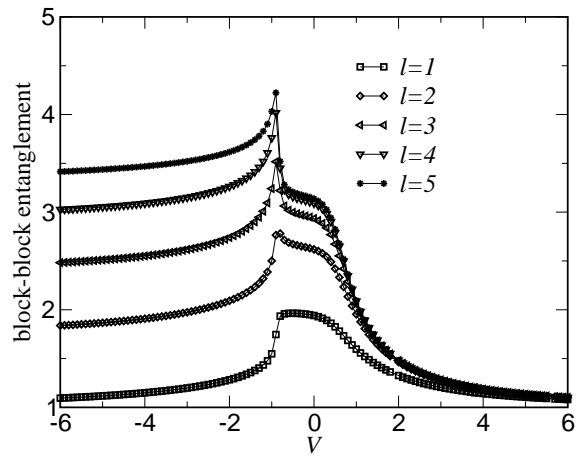}
\caption{The block-block entanglement as a function of $V$  for various block
size. Here $L=10, U=-2$.\label{fig:U=-2} }
\end{figure}

\begin{figure}
\includegraphics[width=7.5cm]{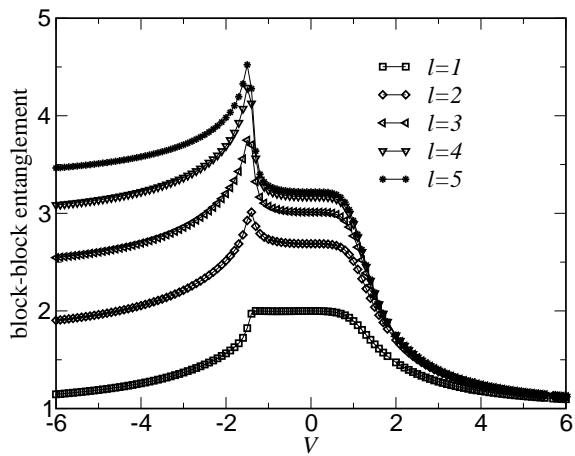}
\caption{The block-block entanglement as a function of $V$  for various block
size. Here $L=10, U=0$. \label{fig:U=0} }
\end{figure}

In Fig. \ref{fig:U=-2}, we show the block-block entanglement as a function
of $V$ for various block size at fixed $U=-2$, $L=10$. In the cross section
of $U=-2$, there are three phases. If $V\ll -1$, the ground state is phase
separated. That is almost half of sites are doubly occupied and congregate
together, while another half of sites are empty. The corresponding
wavefunction is dominated by the following configuration, e.g. for the
10-sites system,
\[{\rm PS(a): } \uparrow\downarrow\hspace{0.25cm}
\uparrow\downarrow\hspace{0.25cm} \uparrow\downarrow\hspace{0.25cm}
\uparrow\downarrow\hspace{0.25cm} \uparrow\downarrow\hspace{0.25cm}
0\hspace{0.25cm} 0\hspace{0.25cm} 0\hspace{0.25cm} 0\hspace{0.25cm} 0, \]
which is in fact an eigenstate of the Hamiltonian with $t=0$. Obviously,
this state is separable and has no entanglement. However, in the presence
of the hoping term, the ground state becomes a superposition of all
possible configurations transformed from the above one through an arbitrary
translation operation. This process introduces entanglement into the ground
state. If we take the reduced-density matrix with block size $l$, the
number of block matrix corresponding to different particle numbers,
$N(l)=0, 2, \cdots, 2l$ is $l+1$. The dimension of the block matrix with a
given nonzero $N(l)=2n, n\le l$, is $l-n+1$, then the number of nonzero
eigenvalues of the reduced density matrix $\rho_l$ is $(l^2+l+2)/2$. This
result implies a $\log(l)$ behavior of the block-block entanglement in the
phase-separated region. That is, the block-block entanglement will not tend
to a constant with the increasing of block size, as shown in Fig.
\ref{fig:U=-2}. This phenomenon is quite different from previous studies of
spin chains \cite{GVidal2003}, where the $\log(l)$ behavior of the
block-block entanglement only exists at the critical region, while at
noncritical region, it will tend to a constant as $l$ increases.

When the absolute value of negative $V$ becomes smaller, the effect of the
hoping term and the on-site interaction can not be neglected. Though the
later still try to maintain the number of doubly occupied sites, the former
really want to diffuse the cumulated electron pairs. Then the configuration
PS(a) will not be the dominant one, and the weights of the following
configurations
$${\rm PS(b): }\uparrow\downarrow\hspace{0.25cm}\uparrow\downarrow\hspace{0.25cm}
  \uparrow\downarrow\hspace{0.25cm}\uparrow\downarrow\hspace{0.25cm}\uparrow\hspace{0.25cm}
\downarrow\hspace{0.25cm}0\hspace{0.25cm}0\hspace{0.25cm}0\hspace{0.25cm}0\hspace{0.25cm}$$
$${\rm PS(c): }\uparrow\downarrow\hspace{0.25cm}\uparrow\downarrow\hspace{0.25cm}
  \uparrow\downarrow\hspace{0.25cm}\uparrow\downarrow\hspace{0.25cm}0\hspace{0.25cm}
\uparrow\downarrow\hspace{0.25cm}0\hspace{0.25cm}0\hspace{0.25cm}0\hspace{0.25cm}0\hspace{0.25cm}$$
will increase. Obviously the configuration PS(b) will introduce block
matrices with odd number of particle into the reduced-density matrix, and
thus increase the entanglement. So the ordered phase is destroyed gradually
as $V$ increases, and finally the system has maximum entanglement and
transits to the SS phase. In the SS phase, the ground state is
characterized by an off-diagonal-long-range order. If $V\rightarrow 0$, the
off-diagonal correlation functions in the reduced-density matrix will cause
a decrease in entanglement which is consistent with the figure. If $V\geq
0$, the hoping process will be suppressed, and the off-diagonal-long-range
order will be destroyed. This fact leads to further decrease of the
block-block entanglement. Moreover, with the increasing of $V$, the weight
of CDW will increase almost to 1, so that the reduced density matrix is
composed of 1/2 $\uparrow\downarrow$ and $0$ each, and the corresponding
entanglement can be calculated exactly:
\begin{eqnarray}
E_v(l)=-\sum_{i=1,2}\frac{1}{2}\log_2{\frac{1}{2}}=1
\end{eqnarray}
This result is consistent with the previous work on the spin model
\cite{GVidal2003}, which says that the block-block entanglement will tend to a
constant in the noncritical region as the block size increases.

In Fig. \ref{fig:U=0}, we show the block-block entanglement as a function
of $V$ for various block size at fixed $U=0, l=10$. Similar analysis for
Fig. \ref{fig:U=-2} can be applied to Fig. \ref{fig:U=0}. The main
difference is that in the nearby region of $V=0$, it seems to be a flat
area. As SDW, CDW, SS and TS all exist in this region, it is not easy to
divide them clearly by using the block-block entanglement only. We also
take more detailed calculation in this region. It is surprising that for
$L=10, U=0$, the block-block entanglement get a minimum value at $V=0.0$,
and a maximum value at $V=0.09$. It seems that the minimum point is a
transition point, just like the appearance of a maximum in the local
entanglement\cite{SJGuPRL} at this point. However, whether these points
witness phase transitions needs further investigation on longer chains.

\begin{figure}
\includegraphics[width=7.5cm]{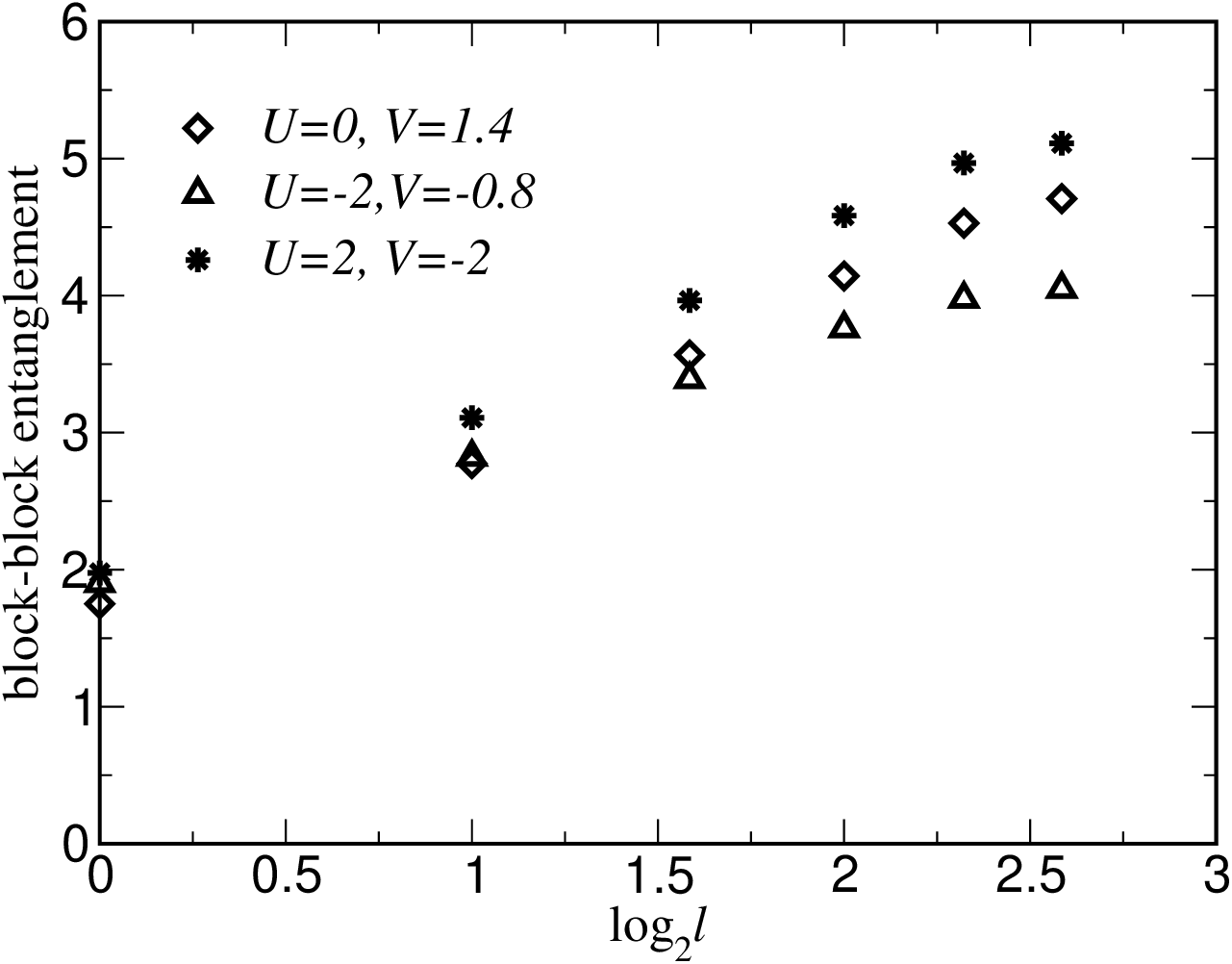}
\caption{The block-block entanglement change with $\log_2{l}$. Here $L=12$.
\label{fig:scaling} }
\end{figure}

>From the above investigations, we find that the block-block entanglement is an
increasing function of $l$ in the region $l \in [0,L/2]$. In our study, in
order to obtain the entropy of reduced density matrix, we need to diagonalize
density matrix by the standard QR algorithm. Limited by the computer power, the
largest size we have studied is 12, then the block size varies from 1 to 6. It
has been argued that the behavior of the entanglement in a critical spin chain
matches the result of the conformal field theory, where the geometry entropy is
analogous to the spin block entropy. Although we can not give the exact result
of scaling on the EHM, we can still see from Fig. \ref{fig:scaling} that the
block-block entanglement grows logarithmically with the block size in some
critical regions. Moreover, as we pointed out in the above, such a logarithmic
behavior of the block-block entanglement can also exist in the noncritical
region for the extended Hubbard model.

\begin{figure}
\includegraphics[width=7.5cm]{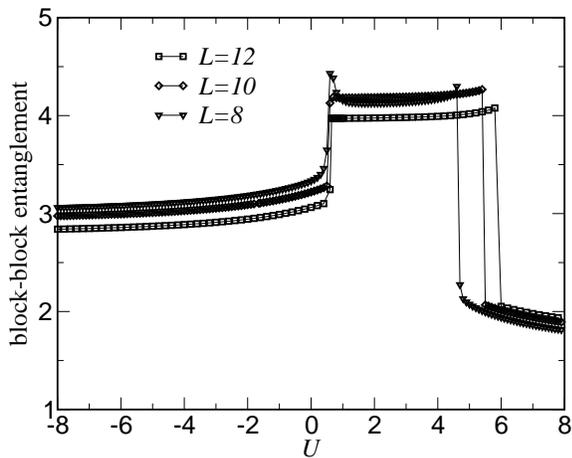}
\caption{The block-block entanglement change with different values of $V$ for
fixed block size $l=4$ and various system size. Here $V=-4$.
\label{fig:scaleforv} }
\end{figure}

%PS(c):\hspace{0.25cm}$\uparrow\downarrow\hspace{0.25cm}\uparrow\downarrow\hspace{0.25cm}
%  \uparrow\downarrow\hspace{0.25cm}\uparrow\hspace{0.25cm}\downarrow\hspace{0.25cm}
%\uparrow\hspace{0.25cm}\downarrow\hspace{0.25cm}
% 0\hspace{0.25cm} 0\hspace{0.25cm} 0$

\section{Entanglement change with system size}
\label{sec:level5}

In this section, we study the dependence of the block-block entanglement on the
system size. In our previous work on the local entanglement of the Hubbard
model, the local entanglement in some regions (CDW and SDW, for example)
obtained from the Bethe Ansatz method (for $L=\infty$ and $L=70$) and by the
exact diagonalization technique ($L=10$) agree with each other excellently.
However, if one wants to study phases such as phase separation, entanglement
will show strong finite size dependence. For simplicity, we look at the local
entanglement again. If $U$ is negative and $V\ll -1$, the ground state is just
a superposition of the configurations transformed from PS(a) by the translation
operation. The components in $\rho_1$ are $z=w\simeq 1/2, u^+=u^-\simeq0$, then
$E_v(1)\simeq 1$. However, if the on-site interaction becomes positive and
larger, the ground-state will prefer the configuration
\[ {\rm PS(d): }0\hspace{0.25cm} 0\hspace{0.25cm} \downarrow\hspace{0.25cm}
\uparrow\downarrow\hspace{0.25cm} \uparrow\downarrow\hspace{0.25cm}
\uparrow\downarrow\hspace{0.25cm} \uparrow\downarrow\hspace{0.25cm}
\uparrow\hspace{0.25cm} 0\hspace{0.25cm} 0, \]
to the PS(a). For the state
dominated by the PS(d) configuration, the local entanglement has a strong
finite size dependence:
\begin{eqnarray}
E_1= \frac{2}{L}\log_2 L-\left( 1-\frac{2}{L} \right) \log_2
\left(\frac{1}{2}-\frac{1}{L}\right).
\end{eqnarray}
Though it will tend to $1$ in the thermodynamic limit, it develops a jump
in the 3D figure of the local entanglement for a small system
\cite{SJGuPRL}. Such a jump in a finite chain signals a crossover from one
PS configurations to another.

We now address the problem for the case of $l>1$. We show the block-block
entanglement change with the value of $V$ for fixed $l=4$ and vary system
size in Fig. \ref{fig:scaleforv}. From the figure, we first notice a jump
around $U=5$, which is the critical point for the transition from PS to
SDW. We also see that there are seemly jumps of the block-block
entanglement around $U=0.6$. However, it is not a critical point. It is
just a result of finite-size effect, as we found in the local entanglement.
Because in both $U < 0.6$ and $0.6 < U < 4$ regions, the ground state is
dominated by phase separation configurations. The difference is that in the
region $U < 0.6$, the dominant configuration is PS(a). As $U$ increases,
one doubly occupied site tends to be singly occupied. So the configuration
PS(d) becomes the dominating configuration. Therefore the jump here does
not correspond to a true critical point and will be suppressed to zero in
the thermodynamic limit.

%%%%%%%%%%%%%%%%%%%%%%%%%%%%%%%%%%%%%%%%%%%%%%%%%%%%%%%%%%%%%%%%%%%%%%%%
% Next time begin here.

\begin{figure}
\includegraphics[width=7.5cm]{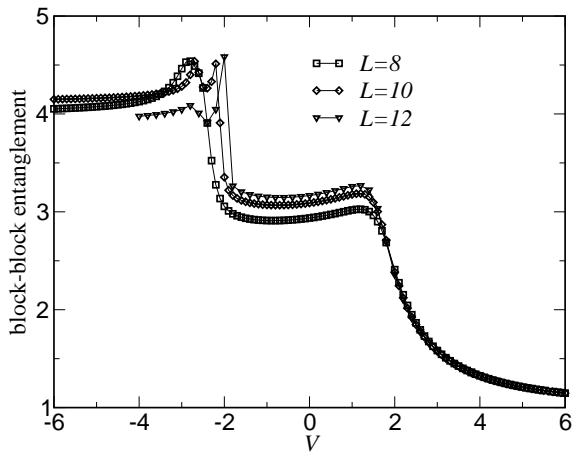}
\caption{The block-block entanglement change with different values of $U$ for
fixed block size $l=4$ and various system size. Here $U=2$.
\label{fig:scaleforu} }
\end{figure}

Another finite-size effect is shown in Fig \ref{fig:scaleforu}. There are two
obviously close apexes for both $L=10$ and $L=12$. For $L=8$, there are two
close apexes for smaller value of $U$. Actually, the first apex is not a
critical point. It is also caused by the crossover between two PSs. Take $L=10$
as an example, for $V < -2.8$, the ground state is dominated by configuration
PS(d), while for $-2.8 < V <-2.2$, it is dominated by configuration PS(a). At
first glance, this result seems unreasonable, because when $V > -2.2$, the
ground state is dominated by SDW. Why the ground state tends to be doubly
occupied before it becomes a SDW phase? A second order perturbation analysis
will help us to understand the complication here. In the strong coupling limit,
$|t/U| \ll 1$ and/or $|t/V| \ll 1$, we have the energy for the configuration
PS(d) and PS(a), respectively, $5U + 16V + 4 t^2/V$ and $4U + 16V + (3/2)
t^2/V$. For small $|U|$, the system prefers to be phase separated due to
attractive Coulomb potential $V$, and the hopping process prefers the
configuration PS(a). Thus the effect of hopping process is stronger than that
of the on-site Coulomb interaction when the absolute values of $V$ decreases,
and the dominating configuration changes from PS(d) to PS(a). Similar to the
argument for Fig. \ref{fig:scaleforv}, this unexpected apex is also a finite
size effect. Because when the block size is fixed, we vary system size to
$\infty$, the change from one pair of separate $\uparrow$ and $\downarrow$ to a
doubly occupied will cause no weight change in the reduced-density matrix.

\section{discussions}
\label{sec:further}

An important observation in our study is that in order to reveal phase
transition by the block-block entanglement, the choice of block type is
essential. Sometimes, the block-block entanglement could be a smoothly
continuous function and shows no obvious structure at the transition point.
As a result, finding the phase boundaries of some transitions is not an
easy job, when the block-block entanglement at the critical point behaves
neither singular nor extremum. Extremal values or discontinuities shown at
some parameters is only suggestive. One must do more detailed analysis to
identity where there exists true QPT and its nature.  A typical example is
the transition from SS to CDW, as shown in Fig. \ref{fig:U=-2}. When such
situation arises, one may ask oneself these questions: (i) What is the
behavior of the block-block entanglement in the critical region, i.e. its
derivatives? (ii) How about choose different block, such as the ``block" in
the momentum space instead of in real space due to off-diagonal-long-range
order?

\begin{figure}
\includegraphics[width=7cm,angle=270]{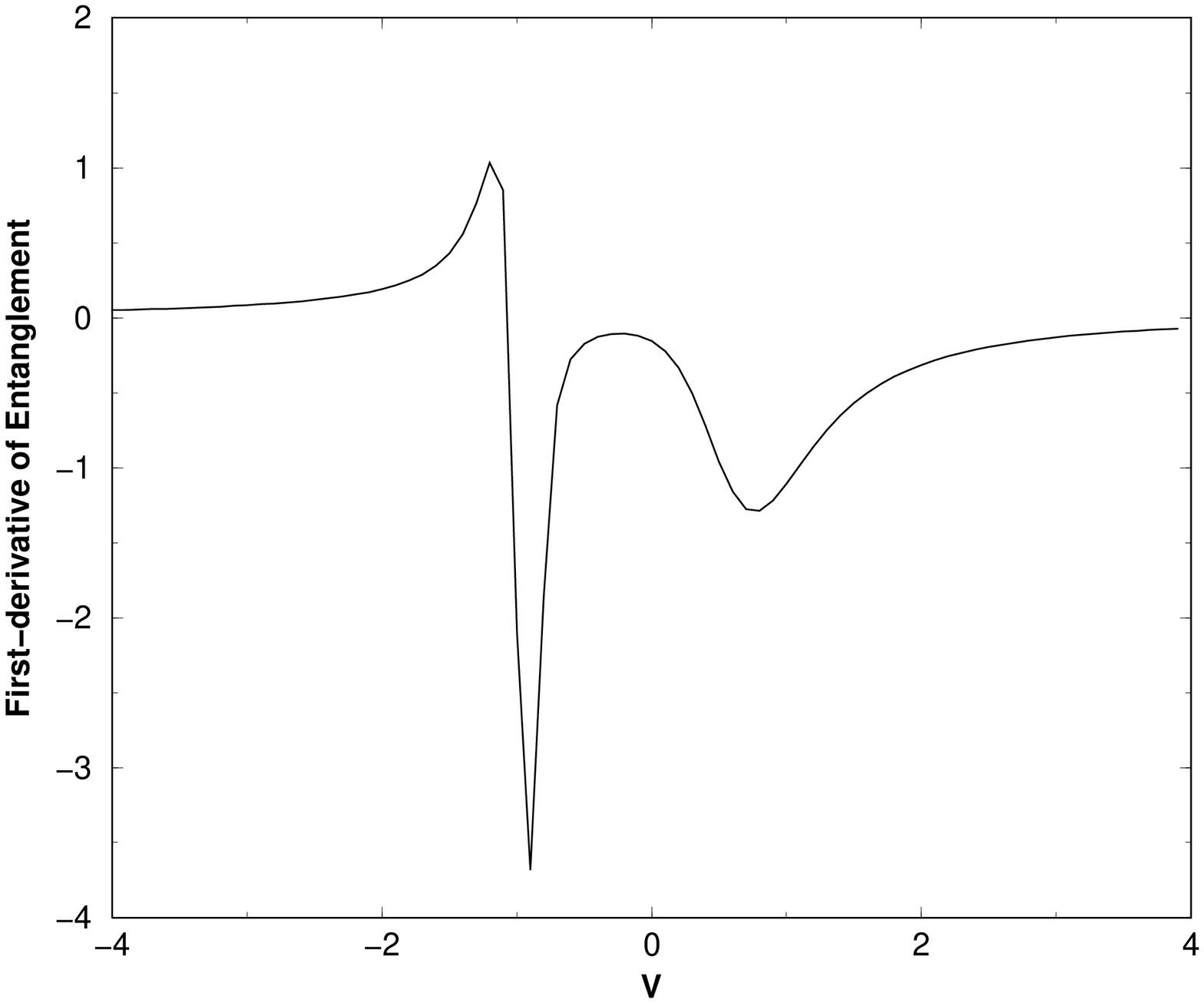}
\caption{The first derivation of block-block entanglement as a
function of $V$. Here $L=8, l=4$ and $U=-2$. \label{fig:deriv8} }
\end{figure}

Regarding the first question, Osterloh {\it et.al.}\cite{AOsterloh2002}
found out that in the one-dimensional transverse-field Ising model, the
first derivative of the concurrence with respect to the coupling diverges
at the critical point, even though the concurrence itself does not show
maximum value at that point. This approach was also applied to other
quantum spin models \cite{JVidal04} and they found similar results. In Fig.
\ref{fig:deriv8}, we show the first derivative of the block-block
entanglement with respect to the nearest-neighbor interaction $V$:
$\partial{E}/\partial{V}$. At the point $V=-0.2$, the derivative of the
block-block entanglement achieves a extreme value, which is very close to
the phase transition point for the infinite chain at $V=0$. Therefore,
besides its extremum behavior in the QPT process, the derivative of the
entanglement may show extremum at critical point.

\begin{figure}
\includegraphics[width=7.5cm]{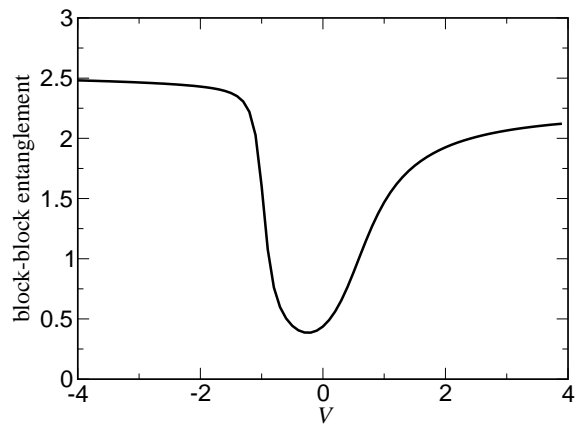}
\caption{The block-block entanglement in momentum space as a function of $V$.
Here $L=8, l=4$, and $U=-2$ \label{fig:kspace} }
\end{figure}

Regarding the second question, we choose the ``block'' in the momentum
space. The choice came from the nature of superconducting paring, for the
order parameter is
\begin{eqnarray}
\triangle_x=\frac{1}{\sqrt{N}}\sum_i
c_{i,\uparrow}^{\dagger}c_{i+x,\downarrow}^{\dagger},
\end{eqnarray}
thus, in the momentum space, one has, for singlet superconducting phase,
\begin{eqnarray}
\triangle_x +
\triangle_{-x}=\frac{2}{N}\sum_k{\cos{kx}c_{k,\uparrow}^{\dagger}c_{{-k},\downarrow}^{\dagger}}
\end{eqnarray}
while for triplet superconducting phase,
\begin{eqnarray}
\triangle_x - \triangle_{-x}=
\frac{2i}{N}\sum_k{\sin{kx}c_{k,\uparrow}^{\dagger}c_{-k,\downarrow}^{\dagger}}.
\end{eqnarray}
So if we choose the block in the momentum space, i.e. $\pm k$, the creating
operator of cooper pairs, which is indeed a superposition of the
superconducting order parameter, is naturally included in the reduced density
matrix.

We compute the block-block entanglement in the momentum space for the $L=8$
system. Using the anti-periodic boundary conditions, the bases we choose
are $K=\pm 1/8\pi, \pm 3/8\pi, \pm 5/8\pi, \pm 7/8\pi$. The results of the
block-block entanglement between $K=\pm 1/8\pi$ and the rest of the momenta
are shown in Fig. \ref{fig:kspace}. From the figure we observe that the
block-block entanglement in the momentum space shows a minimum value at the
point $V=-0.2$. This minimum value may suggest a change of phase from SS to
CDW. Surprisingly, this result is consistent with the result of the first
derivative of the block-block entanglement in real space. However, the
entanglement in the momentum space cannot be used to identify phase
separation in the cross section. This is understandable, because phase
separation occurs in real space.

\section{Summary and acknowledgement}
\label{sec:sum}

In summary, we have studied the block-block entanglement in the ground state of
half-filled one dimensional extended Hubbard model. We found that it can
identify the main phases of EHM on the $U-V$ plane, SDW, CDW, PS and even SS.
Obviously, since the block-block entanglement with block size larger than one
comprises nonlocal correlation, it provides more information than the local
entanglement. Then a richer structure has been obtained from its contour map on
the $U-V$ plane. Moreover, the scaling analysis based on the block size implied
that the ${\rm log}_2(l)$ behavior of the block-block entanglement may also
exist in the noncritical region, such as PS. This result is quite different
from the previous studies of spin systems. We interpret it due to the high
density-of-state closing to the ground state in this region. The number of
different configurations taken part in the ground state is proportional to the
system length. While in other noncritical regions, such as CDW, the ground
state is nearly doubly degenerate and comprises two configurations, this fact
restrict the increase of the block-block entanglement when the block size
increases. By studying the dependence of the entanglement on system length, we
also clarified its unusual behavior in the crossover from one PS to another and
attribute it to the finite size effect. On the other hand, the derivatives of
the block-block entanglement and the block-block entanglement in momentum space
also gave us some interesting results.

However, it seems that the block-block entanglement still can not witness
all phases of the system, such as the BOW and the TS. There are two
possible reasons. One is that the system size we considered is not large
enough, and the other is the limitation of the block-block entanglement
itself. For the former, it is useful to apply other numerical method, such
as density-matrix-renormalization-group, to this problem in the future
studies. While for the later, it is necessary to search other kinds of
proper way to quantify the entanglement in order to have a comprehensive
understanding on the critical behavior in the ground state of a many-body
systems, just as we did in this work. Possible choices include sublattice
entanglement suggested by Chen {\it et al} \cite{YChen04} and the fermion
concurrence proposed by Deng and Gu \cite{SSDeng05}. Obviously, each choice
has its own merits and limitations.

This work is supported by the Earmarked Grant for Research from the Research
Grants Council of HKSAR, China (Project CUHK 401504).


\begin{references}

%EPR
\bibitem{ABinstein35}
A. Einstein, B. Podolsky, and N. Rosen, Phys. Rev. {\bf 47}, 777
(1935).

%Entanglement, Teleportation, Quantum Information
\bibitem{SeeForExample}
See review article by C. H. Bennett and D. P. Divincenzo, Nature
{\bf 404}, 247 (2000).

\bibitem{MANielsenb}
M. A. Nielsen and I. L. Chuang, {\it Quantum Computation and
Quantum Information} (Cambridge University Press, Cambridge,
2000).


\bibitem{YLi05}
Y. Li, T. Shi, B. Chen, Z. Song, and C. P. Sun, Phys. Rev. A {\bf 71}, 022301
(2005).

\bibitem{JZhang05}
J. Zhang, {\it etal}, Phys. Rev. A {\bf 72}, 012331 (2005).


%Entanglement in spin systems
\bibitem{KMOConnor2001}
K. M. O'Connor and W. K. Wootters, Phys. Rev. A {\bf 63}, 052302 (2001).

\bibitem{PZandardi2000}
P. Zanardi, Phys. Rev. A {\bf 65}, 042101(2002).

\bibitem{LFSantos2003}
L.F. Santos, Phys. Rev. A {\bf 67}, 062306 (2003).

\bibitem{XWang2001PLA}
X. Wang, Phys. Lett. A {\bf 281}, 101(2001); X. Wang, and P. Zanardi, Phys.
Lett. A {\bf 301}, 1(2002).

\bibitem{SJGuPRAJ1J2}
S. J. Gu, H. Li, Y. Q. Li, and H. Q. Lin,  Phys. Rev. A {\bf 70}, 052302
(2004).

%Entanglement and QPT
\bibitem{AOsterloh2002}
A. Osterloh, Luigi Amico, G. Falci and Rosario Fazio, Nature {\bf 416}, 608
(2002); T. J. Osborne and M.A. Nielsen, Phys. Rev. A {\bf 66}, 032110(2002)

\bibitem{SJGuXXZ}
S. J. Gu, H. Q. Lin, and Y. Q. Li, Phys. Rev. A {\bf 68}, 042330 (2003); S. J.
Gu, G. S. Tian, H. Q. Lin,  Phys. Rev. A {\bf 71}, 052322 (2005).

\bibitem{JVidal04}
J. Vidal, G. Palacios, and R. Mosseri, Phys. Rev. A {\bf 69}, 022107 (2004); J.
Vidal, R. Mosseri, J. Dukelsky, Phys. Rev. A {\bf 69}, 054101 (2004).

\bibitem{LAWu04}
L. A. Wu, M. S. Sarandy, and D. A. Lidar, arXiv: quant-ph/0407056

\bibitem{MFYang04}
M. F. Yang, arXiv: quant-ph/0407226.

\bibitem{YChen04}
Y. Chen, P. Zanardi, Z. D. Wang, and F. C. Zhang, arXiv: quant-ph/0407228.

\bibitem{FVerstraete04}
F. Verstraete, M. Popp, and J. I. Cirac, Phys. Rev. Lett. {\bf 92}, 027901
(2004); F. Verstraete, M. A. Mart¨ªn-Delgado, and J. I. Cirac, Phys. Rev. Lett.
{\bf 92}, 087201 (2004).

\bibitem{KAudenaert02}
K. Audenaert, J. Eisert, M. B. Plenio, and R. F. Werner, Phys. Rev. A {\bf 66},
042327 (2002).

\bibitem{GVidal2003}
G. Vidal, J. I. Latorre, E. Rico, and A. Kitaev, Phys. Rev. Lett. {\bf 90},
227902 (2003); J. I. Latorre, E. Rico, and G. Vidal, quant-ph/0304098 (2003).

\bibitem{VEKorepin04}
V. E. Korepin Phys. Rev. Lett. {\bf 92}, 096402 (2004).

\bibitem{LViola04}
R. Somma, G. Ortiz, H. Barnum, E. Knill, and L. Viola1, Phys. Rev. A 70, 042311
(2004); H. Barnum, E. Knill, G. Ortiz, R. Somma, and L. Viola, Phys. Rev. Lett.
{\bf 92}, 107902 (2004).


%Quantum Phase Transition: a book
\bibitem{Sachdev}
S. Sachdev, {\it Quantum Phase Transitions}, (Cambridge University
Press, Cambridge, UK, 2000).


%Concurrence definition
\bibitem{Hill} S. Hill and W. K. Wootters, Phys. Rev. Lett. {\bf 78},
5022 (1997); W. K. Wootters, Phys. Rev. Lett. {\bf 80}, 2245 (1998).


%Fermion model
\bibitem{JSchliemann_PRB_63_085311}
J. Schliemann, D. Loss, and A. H. MacDonald, Phys. Rev. B {\bf 63}, 085311
(2001). J. Schliemann {\it et al.} Phys. Rev. A {\bf 64}, 022303 (2001).
%J. Schliemann, J. Ignacio Cirac, M. Ku\'s, M. Lewenstein, and D. Loss,


\bibitem{PZanardi_JPA_35_7947}
P. Zanardi and X. Wang, J. Phys. A: Math. Gen. {\bf 35}, 7947 (2002).

\bibitem{PZanardi_PRA_65_042101}
P. Zanardi, Phys. Rev. A {\bf 65}, 042101 (2002).

\bibitem{SJGuPRL}
S. J. Gu, S. S. Deng, Y. Q. Li, H. Q. Lin, Phys. Rev. Lett. {\bf 93}, 086402
(2004).


\bibitem{JWang03}
J. Wang, and S. Kais, Int. J. Quant. Information {\bf 1}, 375 (2003), and
quant-ph/0405087.

\bibitem{AAnfossi}
A. Anfossi, C. D. E. Boschi, A. Montorsi, and F. Ortolani, cond-mat/0503600.

\bibitem{SSDeng05}
S. S. Deng, S. J. Gu, Chin. Phys. Lett. {\bf 22}, 804 (2005).

\bibitem{DLarsson05}
D. Larsson and H. Johannesson, Phys. Rev. Lett. {\bf 95}, 196406 (2005).


%Quantum phase transition and excited state level crossing
\bibitem{GSTian}
G. S. Tian and H. Q. Lin, Phys. Rev. B {\bf 67}, 245105 (2003).


%Extended Hubbard Model
\bibitem{Emery}
V. J. Emery, pp. 247-303 in {\it Highly Conducting One-Dimensional
Solids}, edited by J. T. Devreese {\it et al.} (Plenum, New York,
1979).

\bibitem{Solyom}
J. Solyom,  Adv. in Phys. {\bf 28}, 201 (1979).

\bibitem{Lin}
H. Q. Lin, Gagliano, D. K. Campbell, E. H. Fradkin, and J. E.
Gubernatis, in {\it The Hubbard Model: Its Physics and
Mathematical Physics}, edited by D. Baeriswyl {\it et al.}, pp.
315-327; see also, H. Q. Lin, D. K. Campbell, and R. T. Clay,
Chin. J. Phys. {\bf 38}, 1(2000).

\bibitem{MNakamura00}
M. Nakamura, Phys. Rev. B {\bf 61}, 16377 (2000).

\bibitem{PSengupta02}
P. Sengupta, A. W. Sandvik, and D. K. Cambbell, Phys. Rev. B {\bf 65}, 155113
(2002); A. W. Sandvik, L. Balents, and D. K. Campbell, Phys. Rev. Lett. {\bf
92}, 236401 (2004).

\bibitem{EJeckelmann02}
E. Jeckelmann, Phys. Rev. Lett. {\bf 89}, 236401 (2002).






%\bibitem{EHLieb68}
%E. H. Lieb and F. Y. Wu, Phys. Rev. Lett. {\bf 20}, 1445 (1968).

%\bibitem{MTakahashib}
%M. Takahashi, {\it Thermodynamics of one-dimensional Solvable
%Models} (Cambridge University Press, Cambridge, 1999).

%\bibitem{ENEconomou79} E. N. Economou and P. N. Poulopoulos, Phys. Rev. B {\bf
%20}, 4756 (1979).

%\bibitem{WMetzner89}
%Walter Metzner and Dieter Vollhardt, Phys. Rev. B {\bf 39}, 4462
%(1989).

%\bibitem{Zhang}
%Y. Chen, P. Zanardi, Z. D. Wang and F. C. Zhang, quant-ph/0407228

%\bibitem{Hamieh}
%S. Hamieh, Acta Physica Polonica B, {\bf 36},3, (2005)

\end{references}
\end{document}